\documentclass[aps,prl,superscriptaddress,preprintnumbers,amsmath,notitlepage,twocolumn,10pt]{revtex4}

\usepackage{graphicx}
\usepackage[squaren]{SIunits}

\begin{document}

\newcommand{\mean}[1]{\left\langle#1\right\rangle}
\newcommand{\fulljust}{\setlength{\rightskip}{0pt}\setlength{\leftskip}{0pt}}
\def\app#1#2{%
  \mathrel{%
    \setbox0=\hbox{$#1\sim$}%
    \setbox2=\hbox{%
      \rlap{\hbox{$#1\propto$}}%
      \lower1.1\ht0\box0%
    }%
    \raise0.25\ht2\box2%
  }%
}
\def\approxprop{\mathpalette\app\relax}

\title{Combined Flux and Anisotropy Searches Improve Sensitivity to Gamma Rays from Dark Matter}

\author{Sheldon S. Campbell}
\affiliation{Center for Cosmology and AstroParticle Physics (CCAPP), Ohio State University, Columbus, OH 43210}
\affiliation{Department of Physics, Ohio State University, Columbus, OH 43210}
\author{John F. Beacom}
\affiliation{Center for Cosmology and AstroParticle Physics (CCAPP), Ohio State University, Columbus, OH 43210}
\affiliation{Department of Physics, Ohio State University, Columbus, OH 43210}
\affiliation{Department of Astronomy, Ohio State University, Columbus, OH 43210}

\date{December 13, 2013}

\begin{abstract}
Searches for dark matter annihilation signals in the diffuse gamma-ray background are promising. We present the first comprehensive study utilizing both flux and anisotropy, using the example of a spectral line signal. Besides improving sensitivity, a combined analysis will help separately determine the particle properties of dark matter and the cosmological aspects of its clumping into halo substructure. The significance of a signal in a shot-noise-dominated anisotropy analysis increases linearly with time $t$, as opposed to $\sqrt{t}$ for a flux analysis, so a flux hint might be confirmed with an anisotropy signal. A first combined line search with Fermi-LAT would provide powerful new sensitivity; one with the proposed GAMMA-400 would be dramatically better.
\end{abstract}

\maketitle

\textbf{Introduction.}---Dark matter's presence in the Universe is inferred from its gravitational effects \cite{Ade:2013zuv}, but its particle nature and small-scale clustering remain unknown. One method for probing dark matter is indirect detection, the search for astrophysical radiation, such as $\gamma$-rays, from dark matter self-annihilation or decay.

Unambiguous identification of dark matter annihilation radiation (DMAR) is challenging, and may require identification from multiple observation channels, such as dwarf galaxies \cite{GeringerSameth:2011iw,Ackermann:2011wa,Ackermann:2013yva}, the Galactic Center \cite{Hooper:2012sr,Gomez-Vargas:2013bea}, neighboring galaxy clusters \cite{Pinzke:2011ek,Han:2012uw}, and the diffuse $\gamma$-ray background (DGRB) \cite{Ackermann:2012rg,Ando:2013ff,Gomez-Vargas:2013cna}. The DGRB is particularly interesting because an observed signal would contain information about both the foreground distribution of dark matter in substructure of the Galactic halo, and throughout the large scale structure of the Universe beyond it. However, DGRB flux searches suffer from the dark matter's unresolved substructure effects being degenerate with its unknown annihilation cross section. Measuring both is required for understanding dark matter \cite{Ng:2013xha}.

We present the first comprehensive sensitivity study of both flux and angular power spectra of the DGRB to dark matter annihilation signals. We illustrate the methods with a spectral line \cite{Mack:2008wu,Essig:2013goa}, but the ideas are more general. We show how these methods can disentangle the dark matter particle properties from substructure effects in the DGRB, how properties of the $\gamma$-ray detector affect signal sensitivity, and surprises in the time dependence of the signals. Each of these is also new.

The fact that $\gamma$-ray anisotropies are potentially sensitive to the presence of annihilating dark matter in Galactic substructure was shown for the case of soft, continuous spectra \cite{SiegalGaskins:2008ge,SiegalGaskins:2009ux,Hensley:2009gh}, where the dark matter signal flux spectrum can be confused with those from astrophysical sources. These works demonstrated how to utilize the angular power spectrum to break that degeneracy.

However, these methods have not yet been applied directly to hard spectra or, more specifically, spectral lines, despite their importance---they are a smoking-gun signature of annihilating or decaying massive particles. A hard spectral feature would appear in both the flux spectrum (due to the source injection spectrum) and fluctuation angular power spectrum (due to the angular distribution of the sources), independently from one another. As has been shown in a different context, the angular signature of $\gamma$-ray lines can be strong \cite{Zhang:2004tj}.

There has been renewed interest in $\gamma$-ray lines with the hint of a signal from the Galactic Center region at an energy near 135 GeV in the Fermi-LAT data \cite{Weniger:2012tx,Bringmann:2012vr,Tempel:2012ey,Su:2012ft}. A recent flux search for $\gamma$-ray lines was completed by the Fermi-LAT collaboration \cite{Fermi-LAT:2013uma}. While the excess near 135 GeV appears statistically interesting, the feature is narrower than the instrument's energy resolution, thus driving the global significance of the signal to below 2$\sigma$. However, an explanation of the feature has not yet been confirmed. More data and investigation will clarify the feature \cite{Weniger:2013tza}, as will improvements to event identification and reconstruction \cite{Atwood:2013rka}. If the line is really from dark matter, the spectral feature will increase in significance, and begin to appear in other regions of the sky. Currently, the significance is decreasing \cite{Weniger:2013tza,Albert}.

In the following, we review the flux sensitivity to spectral lines in statistics-limited experiments, present the first sensitivity analysis of the $\gamma$-ray angular power spectrum to lines, describe the conditions under which the angular power is more sensitive than the flux spectrum, and, as an example, provide the first sensitivity predictions for the tentative 135 GeV line in the DGRB.

\textbf{Number Flux Spectrum.}---We first recall the traditional search of spectral lines with the flux spectrum. 

A high-energy $\gamma$-ray experiment counts detection events, identifying their energy $E$ and direction of origin $\mathbf{\hat n}$. With present experiments currently being statistics limited at high energies \cite{Fermi-LAT:2013uma}, systematic uncertainties are neglected for the sake of this discussion. The number flux in an energy bin $E_i$ from a solid angle $\Omega$ is given by the number $N$ of events via $\Phi=N/(\varepsilon\,\Omega\,t)$, where $\varepsilon$ and $t$ are the instrument's effective area and livetime, respectively.

The line signal is taken to originate from dark matter self-annihilating with cross section $\sigma$. In the Galactic halo, the s-wave annihilation of identical dark matter particles, into a number $N_L$ of $\gamma$-rays per annihilation, produces a spectral line with flux in direction $\mathbf{\hat n}$ of
\begin{equation}
  \Phi_{\text{line}}(\mathbf{\hat n})=\frac{(\sigma v)N_L}{8\pi m^2}J(\mathbf{\hat n}),
\end{equation}
where $v$ is the relative velocity of the annihilating particles and $m$ their mass. The $J$ factor is the line-of-sight integration of the density $\rho$ squared within our halo,
\[
  J(\mathbf{\hat n})=\int ds\ \rho^2(s,\mathbf{\hat n}).
\]
A typical value of mean $J$-factor from the Galactic halo over latitudes $|b|>30^\circ$ is $\mean{J}_{\text{sm}}=\unit{5\times10^{21}}{\giga\electronvolt\squared\usk\centi\meter^{-5}}$, neglecting substructure. The effect of halo substructure is specified by a substructure boost factor $B_\text{sub}$ defined so that $\mean{J}=B_\text{sub}\mean{J}_{\text{sm}}$ over high Galactic latitudes.

Equation~(1) can be generalized to include annihilations beyond the Galactic halo, throughout the cosmos \cite{Ando:2005xg}. These additional signal $\gamma$-rays are important for flux spectrum searches for dark matter \cite{Yuksel:2007ac,Allahverdi:2011sx,Ng:2013xha}, but are neglected to simplify this discussion.

A spectral line is detectable if the signal count $N_{\text{sig}}$ in the energy bin significantly exceeds the uncertainty $\sigma_N$ of the measured total count $N_\gamma$. As long as the experiment is statistics-limited, then $\sigma_N=\sqrt{N_\gamma}$, since $N_\gamma$ is a Poisson statistic. Detection of a line occurs (with a signal size of at least $N_\sigma$ standard deviations) within an energy interval comparable to the instrument's resolution, if the fractional flux over the smooth, diffuse background (e.g., as in \cite{Abdo:2010nz}) is greater than $\Phi_{\text{line}}/\Phi_{\text{tot}}\agt N_\sigma/\sqrt{N_\gamma}$.

\textbf{Fluctuation Angular Power Spectrum.}---We now introduce the line search via the small-scale angular anisotropies of the radiation.

An observable that provides a 2-point correlation of fluctuations on the sky is the \emph{fluctuation} angular power spectrum $\tilde{C}_\ell(E)=\sum_m|\tilde{a}_{\ell m}(E)|^2/(2\ell+1)$, the mean square spherical coefficient at angular partition size $\Delta\theta\sim\pi/\ell$. It is derived from the decomposition of \emph{relative} deviations from the $\mathbf{\hat n}$-averaged flux $\Phi(E_i)$ in each energy bin $E_i$,
\[
  \frac{\Phi(E_i,\mathbf{\hat n})-\Phi(E_i)}{\Phi(E_i)}=\sum_{\ell m}\tilde{a}_{\ell m}(E_i)Y_{\ell m}(\mathbf{\hat n}),
\]
into a basis of spherical harmonics. (Alternatively, \emph{absolute} deviations and the corresponding \emph{intensity} angular power spectrum $C_\ell(E)$ are sometimes used).

The angular power (energy) spectrum of the diffuse $\gamma$-ray background was recently measured by the Fermi-LAT detector \cite{Ackermann:2012uf,Chang:2013ada}. A statistically significant error-weighted mean fluctuation angular power spectrum was found by averaging over multipoles $155\leq\ell\leq504$ in the energy range $\unit{1}{\giga\electronvolt}\leq E\leq\unit{50}{\giga\electronvolt}$ split into 4 energy bins, with an average value of $\tilde{C}_0=\unit{(6.94\pm0.84)\times10^{-6}}{sr}$ after foreground subtraction. The small magnitude of this amplitude is a great opportunity for the detection of even a faint dark matter annihilation signal, because Galactic substructure is currently expected to contribute a much larger fluctuation angular power \cite{Fornasa:2012gu}, as predicted by cosmological simulations \cite{Diemand:2006ik,Diemand:2008in,Springel:2008cc,Stadel:2008pn}.

When considering the sum of different $\gamma$ emitters, each component of the fluctuation angular power spectrum is weighted by the flux ratio squared of the population $j$ of emitters that contributes in energy bin $i$
\begin{equation}
  \label{eq:sumflaps}
  \tilde{C}_\ell(E_i)=\sum_j\left[\frac{\Phi_j(E_i)}{\Phi_{\text{tot}}(E_i)}\right]^2\tilde{C}_{\ell,j}(E_i).
\end{equation}
This follows when the \emph{intensity} angular power spectra superpose linearly, $C_\ell=C_{\ell,1}+C_{\ell,2}+\cdots$, and the fact that $C_\ell(E)=\Phi^2(E)\tilde{C}_\ell(E)$. There will be cross-correlation terms if the population positions are correlated. For this discussion, the main populations are taken as uncorrelated, e.g., Galactic substructure is not correlated with unresolved extragalactic point sources.

We restrict the discussion to Galactic dark matter annihilation, although extragalactic power of the dark matter distribution may be made observable by cross-correlating the $\gamma$-rays with cosmic shear \cite{Camera:2012cj}, or by having very bright annihilation emission in extragalactic halos \cite{Fornasa:2012gu,Ando:2013ff}. Then the $\tilde{C}_{\ell,j}$ associated with dark matter are energy-independent and depend only on the dark matter's spatial distribution (it is the fluctuation angular power of the $J$-factor). Since the contribution from the smooth halo on small angular scales is suppressed in measurements that mask the Galactic plane, the dominant contribution is from halo substructure, $\tilde{C}_{\ell,\text{sub}}$.

Equation~(\ref{eq:sumflaps}) shows the opportunity presented by the small measured value of $\tilde{C}_\ell$ below $\unit{50}{\giga\electronvolt}$. The flux $\Phi_{\text{sub}}$ of the annihilation from substructure does not need to be large to be observable in $\tilde{C}_\ell(E)$. The sensitivity of the angular power spectrum to a spectral line can be determined, similar to the flux spectrum sensitivity.

The expected angular power spectrum from low counts of a purely isotropic source is due to shot noise, well described by $\tilde{C}_N=4\pi f_\text{sky}/N_\gamma$, where $f_\text{sky}$ is the fraction of the sky being observed and $N_\gamma$ is the number of detected $\gamma$-ray events \cite{Ackermann:2012uf}. Angular power in excess of the shot noise is the signal due to source anisotropy. Taking into account beam and mask effects, Fermi-LAT determines its signal angular power $\tilde{C}_\ell$ from the raw measured angular power $\tilde{C}_\ell^{\text{raw}}$ via \footnote{This formula assumes statistical independence of each $C_\ell$ and is valid in multipole bins large enough that neighboring bins are independent. Since we average $C_\ell$ into one large multipole bin, this formula is sufficient for our purposes. See Ref.~\cite{Ackermann:2012uf}, for example.}
\[
  \tilde{C}_\ell^{\text{raw}}=f_{\text{sky}}\left[(W_\ell^{\text{beam}})^2\tilde{C}_\ell+\tilde{C}_N\right],
\]
where the beam window function for a Gaussian beam of beam width $\sigma_b$ is $W_\ell^{\text{beam}}=\exp(-\sigma_b^2\ell^2/2)$. 

As seen in Ref.~\cite{Ackermann:2012uf}, Galactic foregrounds do not significantly affect the angular power for $\ell\agt155$. We assume the data to be statistics-limited at these small angular scales \cite{Ackermann:2012uf}, though a more thorough study of systematics is warranted.

The statistical uncertainty of the measured angular power is \cite{Knox:1995dq,Hinshaw:2003ex,Hinshaw:2006ia}
\begin{align}
  \sigma_{\tilde{C}_\ell}=&\ \sqrt{\frac{2}{(2\ell+1)f_\text{sky}}}\left(\tilde{C}_\ell+\frac{\tilde{C}_N}{(W_\ell^{\text{beam}})^2}\right)\\
  =&\ \tilde{C}_\ell\sqrt{\frac{2}{(2\ell+1)f_\text{sky}}}\left(1+\frac{N_\ell}{N_\gamma}\right),
\end{align}
where
\begin{equation}
  N_\ell\equiv\frac{4\pi f_{\text{sky}}}{\tilde{C}_\ell}e^{\sigma_b^2\ell^2}
\end{equation}
is the approximate number of $\gamma$-rays (for a Gaussian beam) when $\sigma_{\tilde{C}_\ell}$ goes from being shot-noise-dominated, $\propto N_\gamma^{-1}$, to being statistically saturated, $\propto\tilde{C}_\ell$. 

Fermi-LAT will be shot-noise-dominated at high energies throughout the lifetime of the experiment. Thus, the precision of the angular power spectrum improves with time like $N_\gamma^{-1}\propto t^{-1}$, faster than the precision of the flux which improves like $t^{-1/2}$. This is a well-known result in cosmology, not previously noted for dark matter searches, and can be understood by the fact that angular power is a 2-point statistic. When the data are doubled, the number of ways to pair the data is quadrupled, and each pair is statistically independent when the data is noise dominated.

\textbf{Mean-Weighted Angular Power and Line Sensitivity.}---The error-weighted mean of the angular power spectrum over multipoles $\ell_1\leq\ell\leq\ell_2$,
\begin{equation}
  \label{eqn:cmean}
  \tilde{C}\equiv\frac{\sum_{\ell=\ell_1}^{\ell_2}\tilde{C}_\ell/\sigma_{\tilde{C}_\ell}^2}{\sum_{\ell=\ell_1}^{\ell_2}1/\sigma_{\tilde{C}_\ell}^2},
\end{equation}
has variance $\sigma_{\tilde{C}}^2=[\sum \sigma_{\tilde{C}_\ell^2}^{-1}]^{-1}$. In the limit $\ell_2\gg\ell_1$, the sum can be replaced by an integral:
\begin{equation}
  \sigma_{\tilde{C}}=\tilde{C}\sigma_b\sqrt{\frac{2}{f_{\text{sky}}[F_{\ell_2}(N_\gamma)-F_{\ell_1}(N_\gamma)]}},
\end{equation}
with $F_\ell(N)\equiv(1+N_\ell/N)^{-1}-\ln(1+N/N_\ell)$. In addition to the shot-noise-dominated and statistically-saturated regimes, the weighted mean angular power also has a transition zone. These regimes are characterized by
\begin{equation}
  \label{eq:sigCcases}
  \frac{\sigma_{\tilde{C}}}{\tilde{C}}\approx
  \begin{cases}
    \frac{2\sigma_b}{\sqrt{f_{\text{sky}}}}\frac{N_{\ell_1}}{N_\gamma}, & N_\gamma\ll N_{\ell_1}, \\
    \normalsize\vspace{-0.9\baselineskip}\mbox{}\\
    \sigma_b\sqrt{\frac{2}{f_{\text{sky}}[\ln(N_\gamma/N_{\ell_1})-1]}}, & N_{\ell_1}\ll N_\gamma\ll N_{\ell_2}, \\
    \normalsize\vspace{-0.9\baselineskip}\mbox{}\\
    \sqrt{\frac{2}{f_{\text{sky}}(\ell_2^2-\ell_1^2)}}, & N_\gamma\gg N_{\ell_2}.
  \end{cases}
\end{equation}
Consider, for example, the Fermi-LAT analysis of the $\gamma$-ray diffuse background. Applying Eqn.~(\ref{eqn:cmean}) with $\ell_1=155$ and $\ell_2=504$ to the Galactic substructure model in Ref.~\cite{Fornasa:2012gu} (which has $\tilde{C}_{\text{sub},\ell_1}\approx\unit{0.048}{sr}$) yields $\tilde{C}_{\text{sub}}\approx\unit{0.03}{sr}$, over 4000 times larger than the background fluctuations measured by the Fermi-LAT.

We now consider the sensitivity to a \emph{weak} spectral line on a smooth background angular energy spectrum $\tilde{C}_0$ (e.g., extragalactic astrophysical sources). The condition for observing a line signature with a significance of at least $N_\sigma$ is, from Eqn.~(\ref{eq:sumflaps}),
\[
  \tilde{C}-\tilde{C_0}\approx\left(\frac{\Phi_{\text{sub}}}{\Phi_{\text{tot}}}\right)^2\tilde{C}_{\text{sub}}\agt N_\sigma\sigma_{\tilde{C}}.
\]
Therefore, the angular power spectrum will be sensitive to a spectral line from dark matter annihilation if
\begin{equation}
  \frac{\Phi_{\text{sub}}}{\Phi_{\text{tot}}}\agt\sqrt{\frac{N_\sigma\sigma_{\tilde{C}}}{\tilde{C}_{\text{sub}}}}.
\end{equation}
Even with the precision of $\tilde{C}$ improving like $N_\gamma^{-1}$, the sensitivity to the dark matter flux only goes as $N_\gamma^{-1/2}$ (as it does for the flux search).

The condition for the angular power spectrum to be more sensitive to a weak spectral line than the flux spectrum at $N_\sigma$ sensitivity is
\begin{equation}
  \frac{\Phi_{\text{sub}}}{\Phi_{\text{line}}}\agt\sqrt{\frac{N_\gamma\sigma_{\tilde{C}}}{N_\sigma\tilde{C}_{\text{sub}}}},
\end{equation}
before statistical saturation. This shows how the anisotropy analysis becomes more important for higher nominal significances $N_\sigma$. As would be expected, the anisotropy method also becomes more powerful for better angular resolution (small $N_\gamma\sigma_{\tilde{C}}\sim\sigma_b$) and larger $\tilde{C}_{\text{sub}}$.

\textbf{Calculated Sensitivity.}---As a concrete example, consider the tentative Fermi $\gamma$-ray line at 135 GeV, produced by dark matter annihilating to two $\gamma$-rays with cross section $\sigma v=\unit{10^{-27}}{\centi\meter\cubed\usk\reciprocal\second}$. The line flux's nominal significance increases with $t$ as
\[
  N_\sigma=\frac{\Phi_{\text{line}}}{\sigma_\Phi}=\frac{B_{\text{sub}}}{\sqrt{1+(B_{\text{sub}}-1)/F}}\sqrt{R_\Phi t},
\]
where $F\equiv1+\Phi_{\text{BG}}/\Phi_{\text{sm}}$ involves the ratio of the DGRB flux $\Phi_{\text{BG}}$ to the DMAR flux from a smooth Galactic halo profile $\Phi_{\text{sm}}$, and $R_\Phi\equiv \varepsilon\Omega f_{\text{sky}}\Phi_{\text{sm}}/F$. The background flux is calculated in an energy bin of the size of the instrument's energy resolution.

Meanwhile, the fluctuation angular power spectrum's nominal significance can be seen to be
\begin{align}
  N_\sigma&=\frac{|\tilde{C}-\tilde{C}_0|}{\sigma_{\tilde{C}}}=\left|1-\frac{\tilde{C}_0}{\tilde{C}}\right|\sqrt{\frac{f_{\text{sky}}}{2\sigma_b^2}[F_{\ell_2}(N_\gamma)-F_{\ell_1}(N_\gamma)]}\nonumber\\
  &\simeq\frac{\tilde{C}_{\text{sub},\ell_1}}{\unit{0.048}{sr}}\,\frac{(B_{\text{sub}}-1)^2}{1+(B_{\text{sub}}-1)/F}R_C t,
\end{align}
where $R_C=(\unit{0.048}{sr})R_\Phi/(8\pi\sqrt{f_{\text{sky}}}\sigma_be^{\sigma_b^2\ell_1^2})$. This approximation is valid for the most relevant conditions, when $B_{\text{sub}}-1\gg F_{\text{th}}\sqrt{\unit{0.03}{sr}/\tilde{C}_{\text{sub}}}$, where we introduce $F_{\text{th}}\equiv F\sqrt{\tilde{C}_0/\unit{0.03}{sr}}$ (substructure below this threshold is difficult to probe with anisotropy), and $N_\gamma\ll N_{\ell_1}$ (shot-noise-dominated experiment). With $B_\text{sub}$ above the threshold, the time to collect $N_{\ell_1}$ photons is
\begin{equation*}
  t_1\simeq \frac{t_s}{B_\text{sub}-1}\left(1+\frac{F}{B_{\text{sub}}-1}\right)\frac{\unit{0.048}{sr}}{\tilde{C}_{\text{sub},\ell_1}},
\end{equation*}
typically longer than the experiment lifetime for moderate values of $B_{\text{sub}}$, where $t_s\equiv\sqrt{f_{\text{sky}}}/(2\sigma_bFR_C)$.

When the uncertainties are Gaussian deviations, then the nominal significances we defined map to probability confidence intervals in the usual way. In the future, precise confidence intervals can be determined by Monte Carlo simulations, which would also take into account the systematic uncertainties of the detector.

For a future instrument such as GAMMA-400 \cite{Galper:2013sfa}, where the expected angular resolution and energy resolution are both improved from Fermi-LAT by a factor of 10 and the effective area is about $5/8$ smaller, the improved energy resolution affects both methods similarly, but the improved angular resolution gives added benefit to the anisotropy method. Table~\ref{tab:consts} specifies relevant parameters for Fermi-LAT and GAMMA-400.

\begin{table}[!t]
\caption{\label{tab:consts} Constants used for the 135 GeV line sensitivity.}
\begin{ruledtabular}
\begin{tabular}{lccccc}
Experiment & $\sqrt{R_\Phi}~(\text{yr}^{-1/2})$ & $R_C~(\text{yr}^{-1})$ & $F$ & $F_{\text{th}}$ & $t_s~(\text{yr})$\\
\hline
Fermi-LAT & 0.64 & 0.70 & 11 & 0.17 & 21\\
GAMMA-400 & 1.2 & 26 & 2.0 & 0.030 & 30
\end{tabular}
\end{ruledtabular}
\end{table}

\begin{figure}[!b]
  \includegraphics[width=0.23\textwidth]{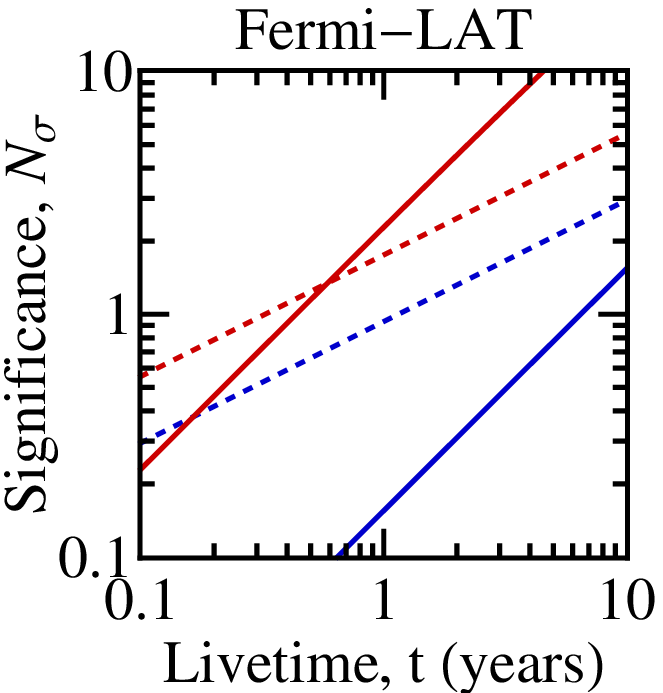}\ \ \includegraphics[width=0.23\textwidth]{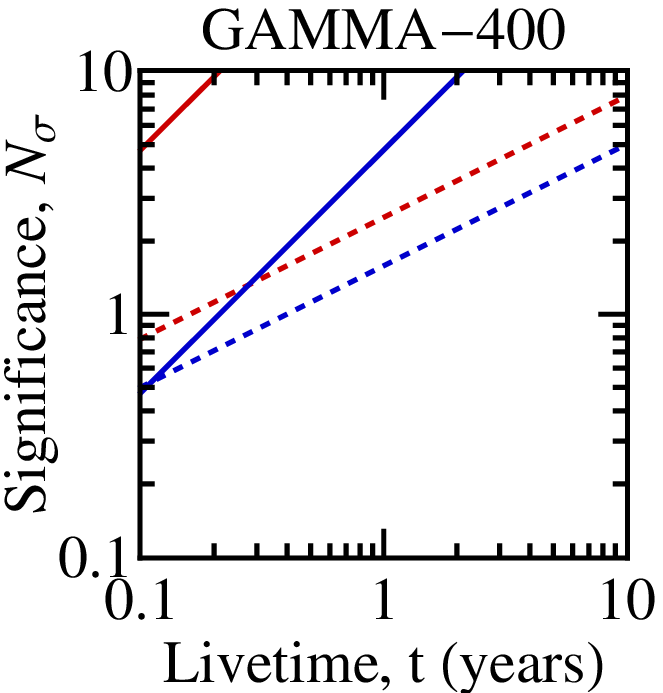}
  \caption{\label{fig:nsig}\fulljust The expected significance (in standard deviations) of the 135 GeV line in the flux spectrum (dashed lines) or the fluctuation angular power spectrum (solid lines) analysis of the diffuse $\gamma$-ray background with the Fermi-LAT or GAMMA-400 experiments, as a function of experiment livetime. We set the cross section to two $\gamma$-rays to be $\sigma v=\unit{10^{-27}}{\centi\meter\cubed\usk\reciprocal\second}$, assumed $\tilde{C}_{\text{sub},\ell_1}=\unit{0.048}{sr}$, and used typical values of $B_{\text{sub}}$ (1.5, lower blue line; 3, upper red line) \cite{Ng:2013xha}. \hfill\mbox{}}
\end{figure}

The significance of the 135 GeV line is shown in Fig.~\ref{fig:nsig} for each experiment, taking $\tilde{C}_{\text{sub},\ell_1}=\unit{0.048}{sr}$ (note that $N_\sigma\propto\tilde{C}_{\text{sub},\ell_1}$ for the anisotropy experiment, when $B_\text{sub}$ is above the threshold). Importantly, the angular signal's significance is not linear with time indefinitely (see Eqn.~(\ref{eq:sigCcases})). For the cases shown here, the lines remain essentially linear for decades. While the flux spectrum may have the first hints of a line at Fermi-LAT, the angular power spectrum is able to verify a discovery more quickly for brighter substructure. For GAMMA-400, the 135 GeV line is generally detected more easily in the angular power spectrum than in the flux spectrum.

\begin{figure}
  \center
  \includegraphics[width=0.36\textwidth]{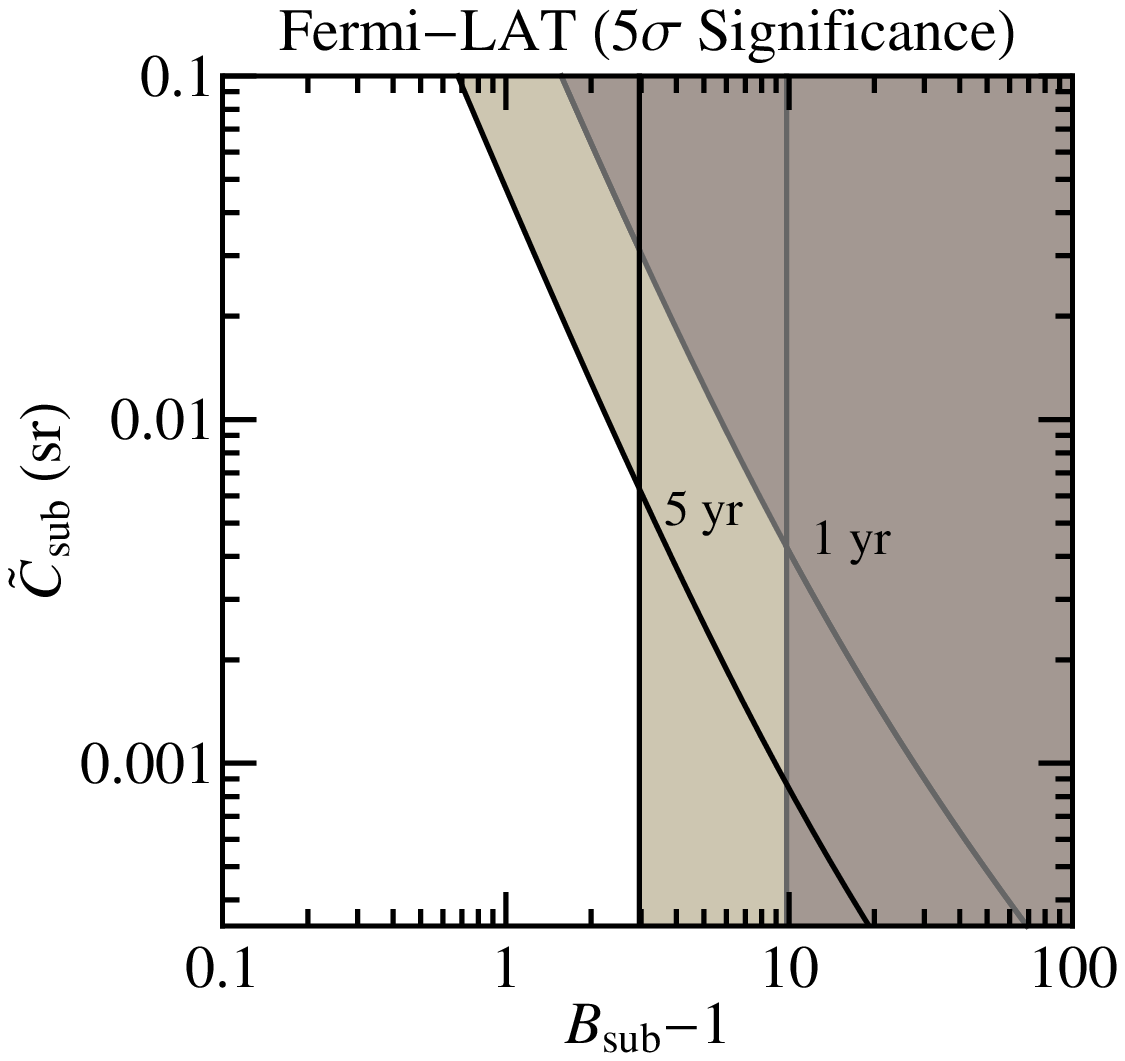}\\
  \vspace{3mm}
  \includegraphics[width=0.36\textwidth]{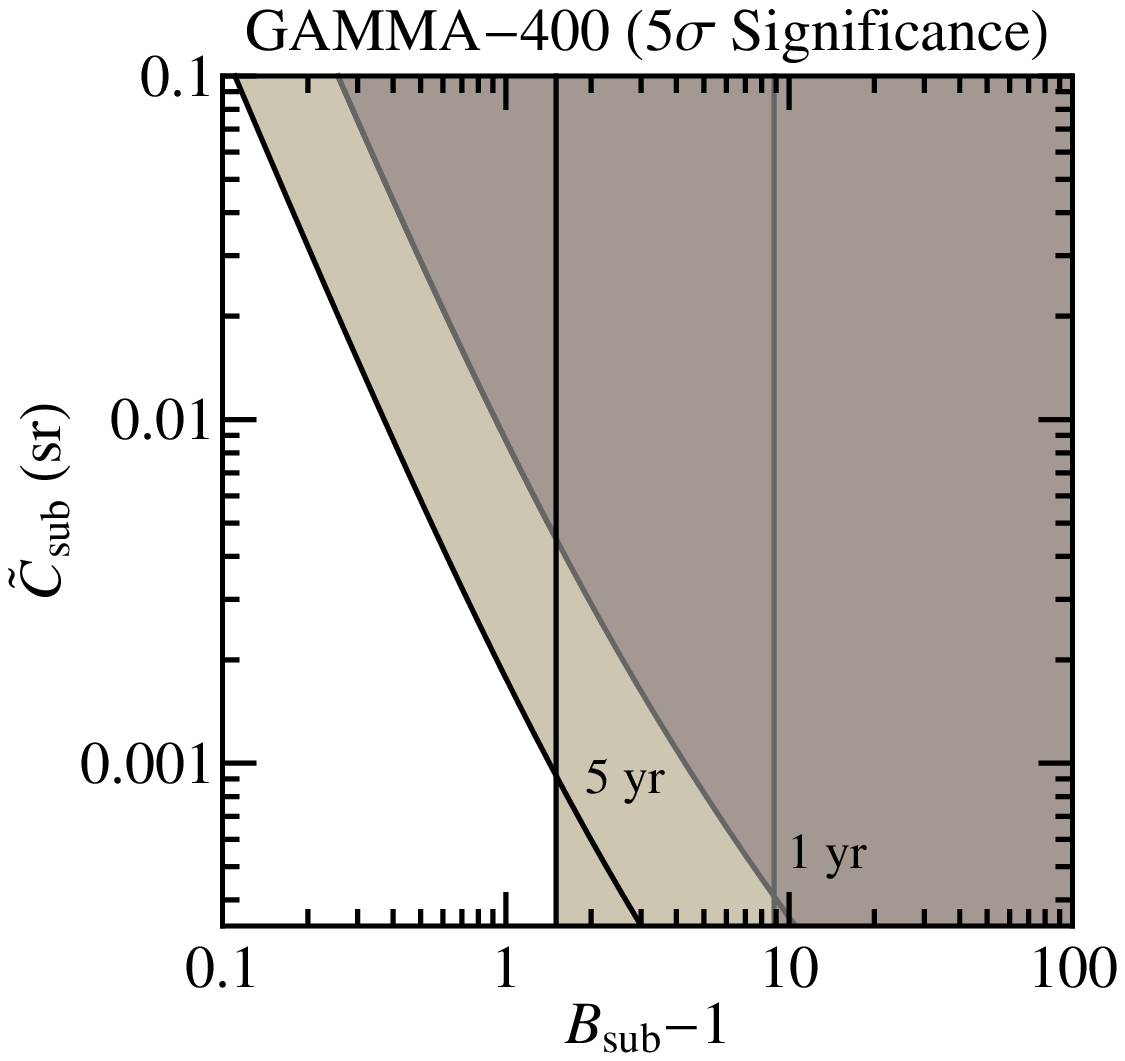}
  \caption{\label{fig:cbplots}\fulljust The parameter space where the 135 GeV line signal has a significance of 5$\sigma$ (a plot at 2$\sigma$ is in Fig.~1 of the Supplementary Material \cite{Supplement}). The vertical lines are the flux sensitivity, while the diagonal curves are the anisotropy. We show the parameter regions where we expect to observe a signal after 1 year (soft lines) or 5 years (dark lines). Note the regions where there would be a signal in one analysis or both. Importantly, realistic dark matter scenarios are being probed and may be discovered with these methods (see text).\hfill\mbox{}}
\end{figure}

Successfully detecting the line in $\tilde{C}$ depends on the values of $B_{\text{sub}}$ and $\tilde{C}_{\text{sub}}$ (which we treat as independent, free parameters) and $\tilde{C}_{\text{sub},\ell_1}$ (which for simplicity we fix to $1.6\tilde{C}_{\text{sub}}$), whereas the flux detection depends only on the value of $B_{\text{sub}}$. The parameter space accessible to each measurement is shown in Fig.~\ref{fig:cbplots} for the Fermi-LAT and GAMMA-400 experiments. The soft lines show the expected limits for observing a 5$\sigma$ signal after 1 year of livetime, whereas the dark lines show the 5-year sensitivity \cite{Supplement}. If $\tilde{C}_{\text{sub}}$ is sufficiently high, the signal becomes visible for smaller values of $B_{\text{sub}}$ than can be detected with flux methods alone. For Fermi-LAT, high $N_\sigma$ detection is more likely to occur first in the anisotropy signal if $\tilde{C}_{\text{sub}}\agt\unit{0.01}{sr}$. In GAMMA-400, the anisotropy signal is important as long as $\tilde{C}_{\text{sub}}\agt\unit{0.001}{sr}$.

In the weak-signal limit ($B_{\text{sub}}\Phi_{\text{sm}}\ll\Phi_{\text{BG}}$), the sensitivity of $B_{\text{sub}}$ in the flux analysis scales for other dark matter lines as $m^2(\sigma v)^{-1}$, whereas the sensitivity of $\tilde{C}_{\text{sub}}$ in the anisotropy analysis scales as $m^4(\sigma v)^{-2}$. Observation of a line in both $\Phi(E)$ and $\tilde{C}_\ell(E)$ fixes the relation between $\tilde{C}_{\text{sub}}$, $\tilde{C}_{\text{sub},\ell_1}$, and $B_{\text{sub}}$, but these parameters may be related in a variety of substructure models, permitting the values $\tilde{C}_{\text{sub}}$ and $B_{\text{sub}}$ to be extracted within particular theoretical substructure frameworks. Knowledge of the line's energy, flux, and $B_{\text{sub}}$ then allows the determination of $\sigma v$ and $m$.

\textbf{Conclusions.}---A combined sensitivity analysis of the flux and angular power energy spectra of a $\gamma$-ray line in the DGRB from annihilating dark matter shows that a combined search increases the sensitivity over using either method alone. Signals in both observables would allow determination of the annihilation cross section and halo substructure properties; both are needed to understand the nature of dark matter \cite{Ng:2013xha}. While flux searches increase in sensitivity like $\sqrt{t}$, angular spectrum signals increase significance more rapidly, like $t$. The presence or absence of the 135 GeV line signal in the 5-year Fermi-LAT DGRB flux and anisotropy data will have interesting consequences for the question of the line's existence. Angular spectrum searches become much more sensitive in a proposed experiment like GAMMA-400. The methodology we presented will generalize to continuum spectra, and most directly to spectra that have most of their detectable signal over background in one energy bin, which is typical.\\
\\
We thank Andrea Albert, Eiichiro Komatsu, Ranjan Laha, and Kenny Ng for discussions and comments on the manuscript. We also thank Eiichiro Komatsu for the explanation of the $N_\gamma^{-1}$ dependence of $\sigma_{\tilde{C}}$. This research was funded by NSF Grant PHY-1101216 to JFB. SC thanks CETUP, supported by DOE Grant DE-SC0010137 and by NSF Grant PHY-1342611, for its hospitality during the 2013 Summer Program where part of this manuscript was produced.

\end{document}